\documentclass[aps,superscriptaddress,nofootinbib,showkeys]{revtex4}%
\usepackage{amsfonts}
\usepackage{amsmath}
\usepackage{amssymb}
\usepackage{graphicx}
\usepackage{hyperref}
\usepackage{bbm}

\def\g{{\mathfrak{g}}}
\def\h{{\mathfrak{h}}}

\newcommand{\be}{\begin{equation}}
\newcommand{\ee}{\end{equation}}
\newcommand{\bea}{\begin{eqnarray}}
\newcommand{\eea}{\end{eqnarray}}

\usepackage[ps,matrix,arrow,curve]{xy}

\usepackage{amsthm}
\theoremstyle{definition}
\newtheorem{theorem}{Theorem}[section]
\newtheorem{lemma}[theorem]{Lemma}

\newtheorem{proposition}[theorem]{Proposition}
\newtheorem{definition}[theorem]{Definition}

\def\nn{\notag}
\def\emph#1{{\sl #1\/}}
\def\Aut{\mathop{\rm Aut}\nolimits}
\def\Der{\mathop{\rm Der}\nolimits}
\def\Lie{\mathop{\rm Lie}\nolimits}
\def\tr{\mathop{\rm tr}\nolimits}
\def\id{\mathop{\rm id}\nolimits}
\def\pullback#1#2{\,{}_{#1}\!\!\times_{#2}}
\def\ker{\mathop{\rm ker}\nolimits}
\def\coker{\mathop{\rm coker}\nolimits}
\def\g{\mathfrak{g}}
\def\h{\mathfrak{h}}

\def\SU{SU}
\def\SO{SO}
\def\Spin{Spin}

\def\eg{{\sl e.g.\/}}
\def\ie{{\sl i.e.\/}}
\def\dd{{\cal D}}

\renewcommand{\1}{\mathbbm{1}}

\begin{document}
\title[Topological Higher Gauge Theory]{Topological Higher Gauge Theory --- from $BF$ to $BFCG$ theory}
\author{\textbf{F. Girelli}}
\email{girelli@sissa.it}
\affiliation{SISSA, 4 via Beirut, Trieste, 34014, Italy and INFN sezione de Trieste}
\author{\textbf{H.\ Pfeiffer}}
\email{pfeiffer@math.ubc.ca}
\affiliation{Department of Mathematics, The University of British Columbia, 1984 Mathematics Road,
             Vancouver, BC, Canada V6T 1Z2}
\author{\textbf{E. M.\ Popescu}}
\email{epopescu@wlu.ca}
\affiliation{Wilfrid Laurier University, Department of Physics and Computer Science,\\
             Waterloo, Ontario, Canada N2L 3C5}
\bigskip
\begin{abstract}

We study generalizations of $3$- and $4$-dimensional $BF$-theory in
the context of higher gauge theory. First, we construct topological
higher gauge theories as discrete state sum models and explain how
they are related to the state sums of Yetter, Mackaay, and Porter.
Under certain conditions, we can present their corresponding continuum
counterparts in terms of classical Lagrangians. We then explain that
two of these models are already familiar from the literature: the
$\Sigma\Phi EA$-model of $3$-dimensional gravity coupled to
topological matter, and also a $4$-dimensional model of
$BF$-theory coupled to topological matter.

\end{abstract}
\date[Date: ]{January 18, 2008}
\maketitle

%
\section{Introduction}
%

Given a $d$-dimensional space-time manifold $M$, a compact Lie group
$G$ with Lie algebra $\g:=\Lie G$, and a principal $G$-bundle
$P\rightarrow M$, the gauge theory associated with the action
\begin{equation}
  S_{BF}(A,B) = \int_M \tr_\g(B\wedge F_A)\label{bfaction}
\end{equation}
is known as $BF$-theory~\cite{Ho89}. Here $F_A$ is the $\g$-valued
curvature $2$-form associated with a connection $A$, $B$ denotes a
$\g$-valued $(d-2)$-form and the notation $\tr_\g(...)$ stands
for the Cartan--Killing form on $\g$.

Although the classical field equations $F_A=0$ and $d_A(B)=0$ are not
very interesting on their own, stating that the connection $A$ be flat
and the $(d-2)$-form $B$ be covariantly constant, $BF$-theory serves
as an important toy model in various contexts.

First of all, the action of $BF$-theory is defined on any smooth
manifold equipped with a principal $G$-bundle, and as such it does not
require the existence of any Riemannian background metric on $M$. This
characteristic is shared with certain first order formulations of
general relativity, \eg\ with the first order formulation of
$3$-dimensional pure Lorentzian [Riemannian] general relativity which
can be shown to be a particular case of $BF$-theory for $d=3$ and with
$G=\SO(1,2)$ or $\Spin(1,2)$ [or with $\SO(3)$ or $\Spin(3)$,
respectively].

Second, $BF$-theory has an enhanced local symmetry, \ie\ besides the
local gauge transformations,
\begin{eqnarray}
  A\mapsto A+\delta A,\qquad\delta A=d_A(\alpha),\\
  B\mapsto B+\delta B,\qquad\delta B=-[\alpha,B],
\end{eqnarray}
where $\alpha$ is locally a $\g$-valued function on $M$, $BF$-theory
is also invariant under the infinitesimal `translations' of the $B$
field,
\begin{equation}
  B\mapsto B+\delta B,\qquad\delta B=d_A(\beta),
\end{equation}
for any $\g$-valued $(d-3)$-form $\beta$.

Finally, since $BF$-theory is independent of any background metric, it
is particularly suitable for the construction of state sum models. For
example, in the case of pure $3$-dimensional Euclidean general
relativity, \ie\ for $3$-dimensional $BF$-theory with $G=\SU(2)$, the
corresponding state sum model is the Ponzano--Regge
model~\cite{PoRe68}. More generally, for an arbitrary compact Lie
group or finite group $G$, one obtains the state sum model by
specializing the Turaev--Viro state sum~\cite{TuVi92,BaWe99} to the
category of finite-dimensional complex representations of $G$ and by
not worrying about convergence of the partition function\footnote{For
the topological interpretation in this case, see, for
example~\cite{BaNa06}.}.

In this article, we study generalizations of $d$-dimensional
$BF$-theory, $d\in\{3,4\}$, to the context of higher gauge
theory~\cite{Ba02,Pf03,GiPf04,BaSch04}.

Roughly speaking, in addition to the connection $1$-form of
conventional gauge theory which equips curves with holonomies in the
gauge group $G$, higher gauge theory introduces a connection $2$-form
which can be used to equip surfaces with a new kind of surface
holonomy, given by elements of another group $H$. More precisely, the
algebraic structure that replaces the gauge group in higher gauge
theory is a crossed module $(G,H,\rhd,t)$, as described in
Section~\ref{sec:xmod} below.

The purpose of the present article is to connect the following
different developments in the literature: (i) higher gauge theory,
\ie\ the generalization of gauge theory from connection $1$-forms to
both $1$-forms and $2$-forms, in the topological case in dimension $3$
and $4$; (ii) state sum invariants of combinatorial $3$- and
$4$-manifolds, familiar from the literature on combinatorial topology,
homotopy theory, and higher category theory; (iii) the $\Sigma\Phi
EA$-model of $3$-dimensional gravity coupled to matter and a related
model of $4$-dimensional $BF$-theory coupled to matter, both of which
are familiar from the literature on quantum gravity.

Our approach to the generalization of $BF$-theory to the framework of higher
gauge theory is therefore twofold. First, we present a combinatorial
construction of such a \emph{topological higher gauge theory} as a state sum
model, and we show that the model is well-defined for any finite crossed
module, \ie\ if the groups $G$ and $H$ are finite. In particular, it makes
sense for arbitrary finite groups, and the group $H$ by which the surfaces are
labeled, is not required to be abelian. We then explain how our state sum
model is related to the state sums of Yetter~\cite{Ye93}, Porter~\cite{Po98},
and Mackaay~\cite{Ma99,Ma00}. In an appendix, we give a self-contained proof
in terms of Pachner moves of the results of Yetter~\cite{Ye93} and
Porter~\cite{Po98,Po96} that these state sums are independent of
triangulation. In fact, both models are invariants of the homotopy type of
$M$~\cite{MaPo06c}. It is an open question which topological invariant
generalizes~\cite{BaNa06} if one studies Lie groups rather than finite groups.

In addition to the combinatorial state sum construction, we present a
continuum counterpart of our models in terms of differential forms and
classical Lagrangians for the case in which $G$ and $H$ are Lie
groups.  At present, two restrictions apply: first, we require the
`fake curvature' (Section~\ref{sec:diffpicture} below) to vanish in
order to make sure that there are well defined curve and surface
holonomies. Second, we require the group $H$ of the crossed module
$(G,H,\rhd,t)$ to be abelian. This is the only case in which the
extended local gauge symmetry is presently fully
understood\footnote{In more general cases, it is not understood
whether one can obtain a Lagrangian that is invariant under a
continuum analogue of the extended local symmetry of~\cite{Pf03} and
whose fields are locally functions or differential forms on
$M$.}~\cite{GiPf04}. There is an obvious candidate for the
corresponding continuum model. We show that one can recover the state
sum by the standard heuristic discretization procedure, and we recall
that special cases of this model have already appeared in the
literature, for example, the $\Sigma\Phi EA$-model~\cite{MaPo06a}.

We emphasize that our state sum models that are available for an
arbitrary finite crossed module $(G,H,\rhd,t)$ yield well-defined
continuum theories, just by considering the continuum limit under
arbitrary refinement of the triangulation. It is, however, not known
whether all such models can be alternatively defined in terms of a
classical Lagrangian.

The present article is structured as follows. In
Section~\ref{sec:preliminaries}, we review the relevant algebraic
tools involved in the description of higher gauge theory: $2$-groups,
crossed modules, Lie $2$-algebras and differential crossed modules. In
Section~\ref{sec:discrete}, we define the discrete state sum models of
topological higher gauge theory in dimensions $d=3,4$. A
self-contained proof that these models are well defined, \ie\
independent of the chosen triangulation, is contained in
Appendix~\ref{app:pachner}. We then explain the relationship to
Mackaay's state sum in Appendix~\ref{app:mackaay}. In
Section~\ref{sec:continuum}, we present the continuum counterparts of
our discrete models for the case of Lie groups and comment on their
relationship with models known from the quantum gravity literature. In
Section~\ref{sec:discretize}, we finally show how the continuum and
discrete models can be related to each other by a generalization of
the usual heuristic discretization procedure.

%
\section{Preliminaries}
%
\label{sec:preliminaries}

\subsection{$2$-Groups}
\label{sec:xmod}

The local symmetry of higher gauge theory is described by an algebraic
structure known as a \emph{categorical group} or as a
\emph{$2$-group}. We first give the relevant definitions and then
briefly sketch in which way this structure is used to equip both
curves and surfaces with holonomies. For more details on $2$-groups
and for a comprehensive list of references, we refer the reader
to~\cite{BaLa04}.

\begin{definition}
A \emph{strict $2$-group} $(G_0,G_1,s,t,\imath,\circ)$ consists of
groups $G_0$ (\emph{group of objects}), $G_1$ (\emph{group of
morphisms}), and homomorphisms of groups $s\colon G_1\to G_0$
(\emph{source}), $t\colon G_1\to G_0$ (\emph{target}), $\imath\colon
G_0\to G_1$ (\emph{identity}) and $\circ\colon
G_1\pullback{s}{t}G_1\to G_1$ (\emph{vertical composition}) such that
the following conditions are satisfied,
\begin{enumerate}
\item
  $s(\imath(g)) = g$ and $t(\imath(g))=g$ for all $g\in G_0$,
\item
  $s(f\circ f^\prime)=s(f^\prime)$ and $t(f\circ f^\prime)=t(f)$
  for all $f,f^\prime\in G_1$ for which $s(f)=t(f^\prime)$,
\item
  $\imath(t(f))\circ f = f$ and $f\circ\imath(s(f))=f$ for all $f\in G_1$,
\item
  $(f\circ f^\prime)\circ f^{\prime\prime}=f\circ(f^\prime\circ
  f^{\prime\prime})$ for all $f,f^\prime,f^{\prime\prime}\in G_1$ for
  which $s(f)=t(f^\prime)$ and $s(f^\prime)=t(f^{\prime\prime})$.
\end{enumerate}
Here $G_1\pullback{s}{t}G_1:=\{\,(f,f^\prime)\in G_1\times
G_1\mid\,s(f)=t(f^\prime)\,\}$ denotes the set of all pairs of vertically
composable morphisms. The multiplication of the groups $G_0$ and $G_1$ is
referred to as \emph{horizontal composition} and is denoted either by
`$\cdot$' or just by simple juxtaposition.
\end{definition}

\begin{definition}
A \emph{strict Lie $2$-group} is a strict $2$-group in which $G_0$ and
$G_1$ are Lie groups and the maps $t,s,\imath$ and $\circ$ are
homomorphisms of Lie groups. A \emph{strict finite $2$-group} is a
strict $2$-group in which both $G_0$ and $G_1$ are finite groups.
\end{definition}

While in ordinary gauge theory, curves are labeled by holonomies
taking values in the gauge group, in higher gauge theory, both curves
and surfaces have holonomies with values in the groups $G_0$ and
$G_1$, respectively:
\begin{equation}
\label{eq_surface}
\begin{aligned}
\xymatrix{
  \bullet\ar@/^2ex/[rr]^{g_1}="g1"\ar@/_2ex/[rr]_{g_2}="g2"&&\bullet
  \ar@{=>}^h "g1"+<0ex,-2ex>;"g2"+<0ex,2ex>
}
\end{aligned}
\end{equation}
The elements $g_1,g_2\in G_0$ label the source and target curves of the
surface, $f\in G_1$ labels the surface, and they are required to satisfy
the conditions,
\begin{equation}
  s(f)=g_1\qquad\mbox{and}\qquad t(f)=g_2.
\end{equation}

The algebraic structure of a strict $2$-group guarantees that one can
change the base points of closed curves and the decomposition of the
boundary of a disc into source and target in a consistent manner and
that one can compose surfaces and define surface-ordered products. In
particular, there is a local gauge symmetry which makes sure that
surface-ordered products are independent of the base point and of the
source curve of the surface~\cite{Pf03,GiPf04}.

Examples of strict $2$-groups can be obtained from Whitehead's crossed
modules of groups as follows.

\begin{definition}
A \emph{crossed module} $(G,H,\rhd,t)$ consists of two groups $G$ and
$H$ and two group homomorphisms $t\colon H\to G$ and $\alpha\colon
G\to\Aut(H),g\mapsto \alpha(g):=(h\mapsto g\rhd h)$, \ie\ an action of
$G$ on $H$ by automorphisms, such that for all $g\in G$ and
$h,h^\prime\in H$,
\begin{eqnarray}
  t(g\rhd h) &=& g\,t(h)g^{-1},\\
  t(h)\rhd h^\prime &=& hh^\prime h^{-1}.
\end{eqnarray}
\end{definition}

\begin{definition}
A \emph{Lie crossed module} is a crossed module in which $G$ and $H$
are Lie groups and in which $t$ and $\alpha$ are homomorphisms of Lie
groups. A \emph{finite crossed module} is a crossed module in which
both $G$ and $H$ are finite groups.
\end{definition}

\begin{proposition}
Given a [Lie, finite] crossed module $(G,H,\rhd,t)$, there exists a
strict [Lie, finite] $2$-group $(G_0,G_1,s,t,\imath,\circ)$ as
follows. The groups of objects and morphisms are $G_0:=G$ and
$G_1:=H\rtimes G$ where the semi-direct product uses the multiplication
$(h_1,g_1)\cdot (h_2,g_2):=(h_1(g_1\rhd h_2),g_1g_2)$. The source and
target maps are given by $s\colon H\rtimes G\to G$, $(h,g)\mapsto g$
and $t\colon H\rtimes G\to G, (h,g)\mapsto t(h)g$, the identity by
$\imath\colon G\to H\rtimes G,g\mapsto (e,g)$ and vertical composition
by $(h,g)\circ (h^\prime,g^\prime)=(hh^\prime,g)$ whenever
$g=t(h^\prime)g^\prime$.
\end{proposition}

In fact, there is a $2$-category of crossed modules and a $2$-category
of strict $2$-groups, and these are equivalent as $2$-categories, see,
for example~\cite{BaLa04,Pf07}. Note that in any strict $2$-group, the
vertical composition is already determined by the remaining structure
maps as $f\circ f^\prime=f\cdot{(\imath(s(f)))}^{-1}\cdot f^\prime$
for all $f,f^\prime\in G_1$ for which $s(f)=t(f^\prime)$, and every
element $f\in G_1$ has got a \emph{vertical inverse}
$f^\times:=\imath(s(f))\cdot f^{-1}\cdot \imath(t(f))$ such that
$f\circ f^\times=\imath(t(f))$ and $f^\times\circ f=\imath(s(f))$.
The map $G_1\to G_1,f\mapsto f^\times$ is a homomorphism of groups.

Using the data of the crossed module, the vertical inverse is
${(h,g)}^\times=(h^{-1},t(h)g)$, $(h,g)\in H\rtimes G$, and the
labeling of the surface of~\eqref{eq_surface} reads,
\begin{equation}
\begin{aligned}
\xymatrix{
  \bullet\ar@/^2ex/[rr]^{g_1}="g1"\ar@/_2ex/[rr]_{g_2}="g2"&&\bullet
  \ar@{=>}^h "g1"+<0ex,-2ex>;"g2"+<0ex,2ex>
}
\end{aligned}
\end{equation}
where $g_1,g_2\in G$ and $h\in H$ are such that $t(h)g_1=g_2$.

\subsection{Lie $2$-algebras}
\label{sec:prelB}

The connection of a conventional gauge theory is often described by using
its connection $1$-form, \ie\ by using a locally defined Lie algebra valued
$1$-form, subject to a certain transformation law under change of the local
trivialization. By analogy, higher gauge theory admits a differential
formulation too, with the role of the Lie algebra of the gauge group being
played by a Lie $2$-algebra. Lie $2$-algebras can be constructed from
differential crossed modules, and in fact, the $2$-category of Lie
$2$-algebras is equivalent as a $2$-category to the $2$-category of
differential crossed modules~\cite{BaCr04}. Here,we just review the
definition and refer to~\cite{BaCr04} for more details and references.

\begin{definition}
A \emph{differential crossed module} $(\g,\h,\rhd,\tau)$ consists of
Lie algebras $\g$ and $\h$ and homomorphisms of Lie algebras
$\tau\colon\h\to\g$ and $d\alpha\colon\g\to\Der(\h), X\to
d\alpha(X):=(Y\mapsto X\rhd Y)$ such that
\begin{eqnarray}
  \tau(X\rhd Y) &=& [X,\tau(Y)],\\
  \tau(Y)\rhd Y^\prime &=& [Y,Y^\prime],
\end{eqnarray}
for all $X\in\g$ and $Y,Y^\prime\in\h$, and with $\Der(\h)$ denoting the
Lie algebra of \emph{derivations} of $\h$.
\end{definition}

Since $d\alpha(X)$ is a derivation of $\h$ for all $X\in\g$, it is
linear and satisfies the relations
\begin{equation}
  X\rhd [Y_1,Y_2] = (d\alpha(X))([Y_1,Y_2])
  = [(d\alpha(X))(Y_1),Y_2] + [Y_1,(d\alpha(X))(Y_2)]
  = [X\rhd Y_1,Y_2] + [Y_1,X\rhd Y_2]
\end{equation}
for all $X\in\g$ and $Y_1,Y_2\in\h$. The map $d\alpha$ is a
homomorphism of Lie algebras, \ie\ it is a linear map that satisfies
the relations
\begin{equation}
  d\alpha([X_1,X_2]) = d\alpha(X_1)\circ d\alpha(X_2) -
   d\alpha(X_2)\circ d\alpha (X_1),
\end{equation}
for all $X_1,X_2\in\g$, \ie
\begin{equation}
  [X_1,X_2]\rhd Y = X_1\rhd (X_2\rhd Y) - X_2\rhd (X_1\rhd Y),
\end{equation}
for all $Y\in\h$. Thus $\rhd$ is an action of $\g$ on $\h$ by
derivations.

\begin{proposition}
\label{prop_diff}
Let $(G,H,\rhd,t)$ be a Lie crossed module. Then there is a
differential crossed module $(\g,\h,\rhd,\tau)$ that can be
constructed as follows. The Lie algebras are $\g:=\Lie G$
and $\h:=\Lie H$, and the homomorphism of Lie algebras
$\tau:=Dt$ is the derivative of the homomorphism of Lie groups
$t\colon H\to G$. If we write $\alpha\colon G\to \Aut(H),
g\mapsto \alpha(g):=(h\mapsto g\rhd h)$, its derivative
$D\alpha\colon\g\to\Der(\h)$ defines the action $\rhd$ in the
differential crossed module by $(d\alpha(X))(Y)=:X\rhd Y$ for all
$X\in\g$ and $Y\in\h$.
\end{proposition}

%
\section{Combinatorial construction of topological higher gauge theory}
%
\label{sec:discrete}

For conventional gauge theory, one can choose the action in such a way
that the theory depends only on the underlying smooth space-time
manifold, but not on any background metric. A very simple example is
given by $BF$-theory~\cite{Ho89} whose classical field equations
require the gauge connection to be flat. In terms of the holonomy
variables, this condition requires the holonomy of any null-homotopic
closed curve to be the identity of the gauge group. We generalize this
idea to the framework of higher gauge theory by imposing the
\emph{higher flatness condition} requiring that the surface holonomy
around the boundary $2$-sphere of any $3$-ball be trivial.

In this section, we present a combinatorial description of such a model for
any triangulation of any smooth manifold of dimension $d\in\{3,4\}$, using the
integral formulation of higher gauge theory~\cite{Pf03}. For $d=3$, this is
precisely the Yetter model~\cite{Ye93} whereas for $d=4$ it coincides with the
Porter's TQFT~\cite{Po98} for $d=4$ and $n=2$. It is known that the partition
function does not depend on the chosen triangulation. In particular, it is
invariant under arbitrary refinement and therefore defines a continuum theory
on the smooth manifold. The renormalization of this model is therefore fully
under control. In fact, we are sitting right on the renormalization fixed
point, and the model is scale invariant. This is no surprise since our
background is just a smooth manifold with no background metric.

The combinatorially defined model is available for any strict finite
$2$-group and even for strict compact Lie $2$-groups if one is not
worried by divergencies similar in nature to those of the Ponzano--Regge
model, \ie\ to those of the $SU(2)$ $BF$-theory in $d=3$~\cite{PoRe68}.

Below, we use the following notation. If $G$ is a finite group with
unit element $e\in G$, we denote by $\int_G dg:=1/|G|\sum_{g\in G}$
the normalized sum over all group elements and by $\delta_G$ the
corresponding $\delta$-distribution on $G$, \ie\ for $g\in G$ we have
$\delta_G(g)=|G|$ if $g=e$ and $\delta_G(g)=0$ if $g\neq e$. If $G$ is
a compact Lie group, $\int_G dg$ and $\delta_G$ denote the Haar
measure and the usual $\delta$-distribution on $G$, respectively.

We define our model for any closed and oriented combinatorial manifold
$\Lambda$ of dimension $d\in\{3,4\}$. These arise precisely as the
triangulations of closed and oriented smooth manifolds of dimension
$d$~\cite{Wh40,Ce68}\footnote{For the relevance of the latter
reference, see~\cite{Ku62}.}. We denote the set of all $k$-simplices,
$0\leq k\leq d$, by $\Lambda_k$. We equip the set of vertices
$\Lambda_0$ which can be assumed to be finite, with an arbitrary total
order and denote the $k$-simplices by $(k+1)$-tuples of vertices
$(i_0\ldots i_k)$ where $i_0,\ldots,i_k\in\Lambda_0$ such that
$i_0<\cdots<i_k$.

\begin{definition}
\label{def_statesum}
Let $\Lambda$ be a compact and oriented combinatorial $d$-manifold,
$d\in\{3,4\}$, and $(G,H,\rhd,t)$ be a finite crossed module. The partition
function of \emph{topological higher gauge theory} is defined by
\begin{eqnarray}
\label{eq_partition}
  Z &=& {|G|}^{-|\Lambda_0|+|\Lambda_1|-|\Lambda_2|}{|H|}^{|\Lambda_0|-|\Lambda_1|+|\Lambda_2|-|\Lambda_3|}\,
        \biggl(\prod_{(jk)\in\Lambda_1}\int\limits_G dg_{jk}\biggr)\,
        \biggl(\prod_{(jk\ell)\in\Lambda_2}\int\limits_H dh_{jk\ell}\biggr)\nn\\
    &\times&
        \biggl(\prod_{(jk\ell)\in\Lambda_2}\delta_G\bigl(t(h_{jk\ell})g_{jk}g_{k\ell}g_{j\ell}^{-1}\bigr)\biggr)
        \biggl(\prod_{(jk\ell m)\in\Lambda_3}\delta_H\bigl(h_{j\ell m}h_{jk\ell}(g_{jk}\rhd h_{k\ell m}^{-1})h_{jkm}^{-1}\bigr)\biggr).
\end{eqnarray}
\end{definition}

Here we integrate over $g_{jk}\in G$ for every edge $(jk)\in\Lambda_1$
and over $h_{jk\ell}\in H$ for every triangle
$(jk\ell)\in\Lambda_2$. The $\delta$-distributions unter the integral
impose the condition that $t(h_{jk\ell})g_{jk}g_{k\ell}=g_{j\ell}$
for each triangle $(jk\ell)\in\Lambda_2$, \ie\ that each surface label
$h_{jk\ell}$ has got the appropriate source and target,
\begin{equation}
\label{eq_triangle0} \xymatrix{
  &&\ell\\
  \\
  j\ar[uurr]^{g_{j\ell}}="t"\ar[rr]_{g_{jk}}&&k\ar[uu]_{g_{k\ell}}
  \ar@{=>}_{h_{jk\ell}} "3,3"+<-2.5ex,2.5ex>;"t"+<2.5ex,-2.5ex>
}
\end{equation}
and the condition that the surface holonomy around every tetrahedron
$(jk\ell m)\in\Lambda_3$ be trivial. Recall from~\cite{GiPf04} that
the argument of the $\delta_H$ in~\eqref{eq_partition} is precisely
the surface ordered product around the tetrahedron:
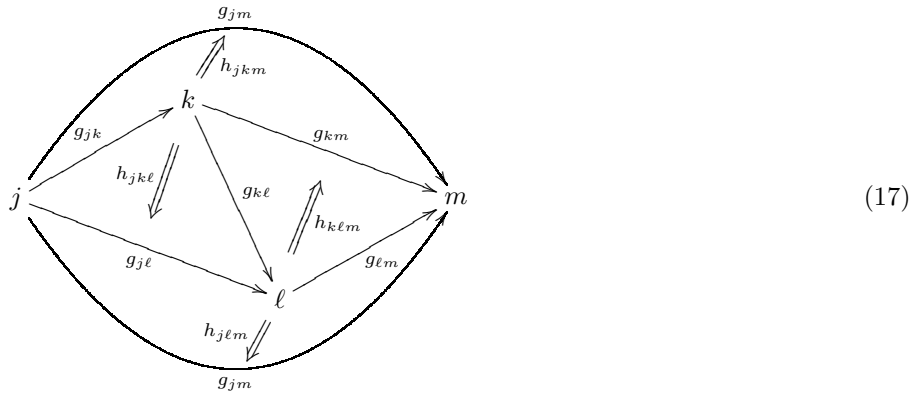
\begin{equation}
\label{eq_tetrahedron0}
\begin{aligned}
\xymatrix{
  &&k\ar[drrr]^{g_{km}}="d"\ar[ddr]^{g_{k\ell}}\\
  j\ar@/^15ex/[rrrrr]^{g_{jm}}="a"\ar@/_15ex/[rrrrr]_{g_{jm}}="b"\ar[drrr]_{g_{j\ell}}="c"\ar[urr]^{g_{jk}}&&&&&m\\
  &&&\ell\ar[urr]_{g_{\ell m}}
  \ar@{=>}_{h_{jkm}} "1,3"+<1ex,2ex>;"a"+<-1ex,-2ex>
  \ar@{=>}_{h_{j\ell m}} "3,4"+<-1ex,-2ex>;"b"+<1ex,2ex>
  \ar@{=>}_{h_{jk\ell}} "1,3"+<-1ex,-4ex>;"c"+<1ex,4ex>
  \ar@{=>}_{h_{k\ell m}} "3,4"+<1ex,4ex>;"d"+<-1ex,-4ex>
}
\end{aligned}
\end{equation}
This expression is independent of the choice of the base edge $(jm)$
because $\delta_H$ is a gauge invariant
function~\cite{Pf03}. Similarly, exploiting the local gauge symmetry
of higher gauge theory~\cite{Pf03,GiPf04}, it is not difficult to show
that the partition function~\eqref{eq_partition} does not depend on
the ordering of the vertices.

Using Alexander moves~\cite{Al30}, Yetter~\cite{Ye93} has shown for
the case $d=3$ that the partition function does not depend on the
chosen triangulation. It seems to have gone unnoticed that the same
result for $d=4$ is in fact already implied by~\cite{Po98} in
combination with~\cite{Po96}, again by using Alexander moves:

\begin{theorem}
Let $\Lambda$ be a closed and oriented combinatorial $d$-manifold,
$d\in\{3,4\}$, and $(G,H,\rhd,t)$ be a finite crossed module. The
partition function~\eqref{eq_partition} is invariant under Pachner
moves and therefore well defined on equivalence classes of
combinatorial manifolds.
\end{theorem}

Since the original references may not be very accessible to readers
interested in higher gauge theory, we sketch in
Appendix~\ref{app:pachner} how one can obtain a self contained proof
of triangulation independence using Pachner moves~\cite{Pa91}. This
has the advantage that there are only a finite number of moves to
verify -- a number that is independent of the chosen triangulation --
and that it can be done in a direct calculation without any additional
machinery from homotopy theory.  In Appendix~\ref{app:mackaay}, we
explain under which conditions the model~\eqref{eq_partition} forms a
special case of Mackaay's state sum~\cite{Ma99,Ma00} and in which
cases it does not.

In the partition function~\eqref{eq_partition}, the labeling of edges
by elements $g_{jk}\in G$ and of triangles with elements
$h_{jk\ell}\in H$ are called \emph{colorings}. Those colorings for
which the $\delta_G(...)$ and $\delta_H(...)$ are non-zero, are called
\emph{admissible colorings}. The partition
function~\eqref{eq_partition} counts the number of admissible
colorings and multiplies the result by
${|G|}^{-|\Lambda_0|}{|H|}^{|\Lambda_0|-|\Lambda_1|}$. This factor may
indicate that one has already integrated out further variables
associated with the lower-dimensional simplices, see, for
example~\cite{LaPf06}. The partition function~\eqref{eq_partition} is
known to be an invariant of the homotopy type of
$\Lambda$~\cite{MaPo06c}.

We emphasize that because of~\cite{Wh40,Ce68}, the
model~\eqref{eq_partition} which we have here defined in the discrete
language of combinatorial manifolds, is in fact a proper continuum
theory that is well-defined on any smooth $d$-manifold. This argument
can be put in a more physical language by saying that the invariance
under the $1\leftrightarrow(d+1)$ Pachner move~\cite{Pa91} allows us
to pass to an arbitrary refinement of the triangulation and thereby to
the continuum limit of the model.

%
\section{Lagrangian formulation}
%
\label{sec:continuum}

If the state sum model (Definition~\ref{def_statesum}) is studied for
a Lie crossed module $(G,H,\rhd,t)$ rather than a finite crossed
module, the partition function~\eqref{eq_partition} is in general no
longer well defined. Work~\cite{FrLo03,BaNa06} on the Ponzano--Regge
model, \ie\ the $H=\{e\}$, $d=3$ special case, nevertheless indicates
that there are physical observables that can still be defined. The
analogy with the Ponzano--Regge model also suggests that in the Lie
group case, there is an alternative, continuous, formulation of the
model in terms of a classical Lagrangian and fields given locally by
differential forms on $M$. For the Ponzano--Regge model, this turned
out to be $SU(2)$ $BF$-theory in $d=3$. In the following, we present a
similar continuum counterpart of the state sum
model~\eqref{eq_partition}.

\subsection{Higher gauge theory in the differential formulation}
\label{sec:diffpicture}

In this section, we recall the differential formulation of higher gauge
theory~\cite{GiPf04} for the case in which the structure $2$-group is given by
a Lie crossed module $(G,H,\rhd,t)$ with $H$ abelian. If we consider the
associated differential crossed module $(\g,\h,\rhd,\tau)$ (see
Proposition~\ref{prop_diff}), the connections of the higher gauge theory
formalism will be a $\g$-valued \emph{connection $1$-form} $A$ and an
$\h$-valued \emph{connection $2$-form} $\Sigma$. Here, we consider only the
local description of the connection of higher gauge theory. For the global
aspects, we refer to~\cite{BaSch04}. Note that our notation is different
from~\cite{GiPf04} where the connection $2$-form was called $B$ rather than
$\Sigma$. In the following, we use the letter $B$ for a further field.

The curvature of higher gauge theory is then given by two differential
forms: the curvature $2$-form ('fake curvature')
\begin{equation}
  F_A = R_A + \tau(\Sigma),
\end{equation}
where $R_A = dA + \frac{1}{2}[A,A]$ is the conventional curvature of
$A$, and the curvature $3$-form
\begin{equation}
  G_\Sigma = d_A(\Sigma) := d\Sigma + A\rhd\Sigma.
\end{equation}

If the connection $1$- and $2$-forms originate from an integral formulation in
terms of holonomies, the fake curvature vanishes: $F_A=0$. In the model that
we introduce in the following section, this condition is enforced
on-shell. Together with the requirement that $H$ be abelian, it ensures that
the model has the following extended local gauge symmetry:
\begin{eqnarray}
  A\mapsto A+\delta A,\quad &\mbox{where}&
    \quad\delta A=d_A(\alpha) + \tau(\lambda)\label{eqn_gsym},\\
\label{gaugetransflagrange}  
  \Sigma\mapsto\Sigma+\delta\Sigma,\quad&\mbox{where}&
    \quad\delta\Sigma=-d_A(\lambda)-\alpha\rhd\Sigma,
\end{eqnarray}
where the gauge transformation is generated locally by a $\g$-valued $0$-form
$\alpha$ and an $\h$-valued $1$-form $\lambda$. The fake curvature and the
curvature $3$-form transform as follows,
\begin{eqnarray}
\label{eq_transformf}
  F_A\mapsto F_A+\delta F_A,\quad &\mbox{where}&
    \quad\delta F_A=[F,\alpha],\\
\label{eq_transformg}
  G_\Sigma\mapsto G_\Sigma+\delta G_\Sigma,\quad &\mbox{where}&
    \quad\delta G_\Sigma=-\alpha\rhd G_\Sigma - F_A\rhd\lambda.
\end{eqnarray}
For more details, the reader is referred to~\cite{Ba02,GiPf04,BaSch04}.

We are in particular interested in the Lie crossed module $(G,H,\rhd,t)$
associated with the adjoint $2$-group of a Lie group $G$. Here $\rhd$ is the
adjoint action of $G$ on its Lie algebra $H:=\g=\Lie G$, and $t(h)=e$ for all
$h\in H$. Its group of morphisms $\g\rtimes G$ is often called the
inhomogeneous group associated with $G$, see, for example~\cite{ashtekar}.

The corresponding differential crossed module (Proposition~\ref{prop_diff}) is
given by $(\g,\h,\rhd,\tau)$ where $\h$ is the vector space underlying $\g$
equiped with the abelian Lie algebra structure, $\g$ acts on $\h$ by the
adjoint action, and $\tau(Y)=0$ for all $Y\in\h$. The corresponding Lie
$2$-algebra has the semidirect sum $\h\oplus_\lhd\g$ as its Lie algebra of
morphisms, the inhomogeneous algebra associated with $\g$.

With this choice of differential crossed module, the vanishing of
the fake curvature $F_{A}$ implies the vanishing of the conventional
curvature $R_A=0$.

\subsection{The $BFCG$ theory}
\label{sec:BFCG}

With the above considerations, we can now propose a classical action
corresponding to the partition function~\eqref{eq_partition} as
follows. Let $(G,H,\rhd,t)$ be the Lie crossed module associated with
the adjoint $2$-group of the Lie group $G$ and $(\g,\h,\rhd,\tau)$ be
the associated differential crossed module as explained above.

Besides the curvature $1$- and $2$-forms of higher gauge theory, we
consider two additional fields $B$ and $C$ which are a $\g$-valued
$(d-2)$-form and an $\h$-valued $(d-3)$-form, respectively, and which
are assumed to transform under the 2-gauge transformations in
(\ref{eqn_gsym}) as:
\begin{eqnarray}
\label{gaugetransflagrangebis}  
  B\mapsto B+\delta B&\rm{with}& \delta B= [B,\alpha]-[C,\lambda],\\
\label{eq_transformc}
  C\mapsto C+\delta C&\rm{with}& \delta C= -\alpha\rhd C.
\end{eqnarray}
It should be noted that with the above choice of differential crossed module,
the local symmetries of both fields $B$ and $C$ in (\ref{gaugetransflagrange})
are well defined and make sense.

The reason for introducing these fields is the same as in $BF$-theory:
their associated field equations are the conditions that the curvature
$2$-form $F_A$ and the curvature $3$-form $G_\Sigma$ vanish. The
action of our $BFCG$ theory therefore reads,
\begin{equation}
\label{action}
  S=\int_M\tr_\g(B\wedge F_A\big)+\tr_\h(C\wedge G_\Sigma).
\end{equation}
Note that the transformations~\eqref{gaugetransflagrangebis}
and~\eqref{eq_transformc} are chosen such as to make the action gauge
invariant in view of the transformations~\eqref{eq_transformf}
and~\eqref{eq_transformg}.

Similar to the traditional $BF$ action in (\ref{bfaction}), the $BFCG$
action~\eqref{action} also exhibits an extended local symmetry, in the
sense that it is also invariant under the additional infinitesimal
gauge transformations:
\begin{eqnarray}
  B\mapsto B+\delta' B&\rm{with}& \delta' B=d_{A}(\beta)+[\Sigma,\gamma],\nn\\
  C\mapsto C+\delta' C&\rm{with}& \delta' C= d_{A}(\gamma),
\end{eqnarray}
where $\beta$ and $\gamma$ are locally a $\g$-valued $(d-3)$-form and
an $\h$-valued $(d-4)$-form, respectively. It should be emphasized
that not all of the above gauge transformations of the fields of the
theory are irreducible. Indeed, on shell, the gauge transformations
for the connection $2$-form $\Sigma$ in (\ref{gaugetransflagrange}),
and in the case $d=4$ also the gauge transformation for the field $B$
in (\ref{gaugetransflagrangebis}), are themselves invariant under an
infinitesimal translation of the gauge parameters
\begin{eqnarray}
  \lambda\mapsto\lambda+\delta' \lambda&\rm{with}& \delta' \lambda=d_{A}(\rho),\nn\\
  \beta\mapsto \beta+\delta' \beta&\rm{with}&\delta' \beta= d_{A}(\eta),\label{redux_sym}
\end{eqnarray}
where $\eta$ and $\rho$ are $\h$-valued and $\g$-valued 0-forms,
respectively. While mathematically obvious, the transformations in
(\ref{redux_sym}) also have a rather straightforward physical
interpretation. A not so complicated counting argument~\cite{MaPo06a}
shows that if the above gauge symmetry reducibility is ignored, in
both the $d=3$ and $d=4$ cases $BFCG$ theory would exhibit a negative
number of (local) physical degrees of freedom. The role of the
transformations in (\ref{redux_sym}) is to bring the number of (local)
physical degrees of freedom up to zero and hence establish the
topological character of the theory.

Upon first order variation, the $BFCG$ action  yields the equations of motion:
\bea
  d_{A}(B)+[\Sigma,C]&=&0\nonumber\\
  d_{A}(C)&=&0\nonumber \\
  F_A&=&0\nonumber\\
  G_{\Sigma}&=&0
\label{eqn21}
\eea
so that in particular the higher flatness condition $G_\Sigma=0$ holds.
Note that the vanishing of the fake curvature $F_A=0$ is
automatically satisfied on-shell.

In dimensions $d=3$ and $d=4$, the $BFCG$-theory can readily be related
with topological models that have already been studied in the literature.
For $d=3$, the $BFCG$ model can be related to Euclidean and
Lorentzian gravity. Indeed if one chooses $\g$ to be the Lie algebra
$so(3)$ or $so(2,1)$ and $\h$ to be the abelian Lie algebra of
3-dimensional translations $t^{3}$, the 1-form field $B$ can be
interpreted as the local triad field of the spacetime manifold $M$,
and the first term in~\eqref{action} becomes the action for pure
gravity in the Palatini formalism. Under these circumstances, and upon
tracing, the $BFCG$ action becomes functionally identical to the
action of a topological matter model that has already been studied in
the literature~\cite{MaPo06a}, called the $\Sigma\Phi EA$
model. Within this latter context, the second term of the $BFCG$
action containing the 0-form field $C$ and the curvature $G_\Sigma$ of
the connection $2$-form $\Sigma$ can be interpreted as a coupling of
topological matter fields to pure 3-dimensional gravity. Consequently,
the $BFCG$ model can be shown to admit topological solutions like
point-particle solutions, the BTZ black-hole solution and cosmological
solutions of the Robertson-Friedman-Walker type~\cite{MaPo06b}.

For $d=4$, by choosing $\g$ to be the Lie algebra $so(4)$ or $so(3,1)$
and $\h$ the $6$-dimensional abelian Lie algebra, it can be shown ---
in a manner similar to the 3-dimensional case~\cite{MaPo06b} --- that
the $BFCG$ action yields (up to surface terms) yet another topological
matter model that has been studied previously in the
literature~\cite{Bi93}. This topological matter model is also
non-trivial~\cite{BiGe94} and is related to topological gravity in
4-dimensional spacetimes in a similar way as ordinary $BF$ theory.

Also note that dropping the second term of~\eqref{action} does not
specialize $BFCG$-theory to the first example in Section~3.9
of~\cite{GiPf04}.

%
\section{Discretization}
%
\label{sec:discretize}

In this section, we show how the action~\eqref{action} is related to
the state sum model~\eqref{eq_partition} by the usual heuristic
discretization procedure. We therefore consider the partition function
\begin{equation}
\label{eq_partcont1}
  Z = \int [\dd C][\dd B][\dd A][\dd\Sigma] e^{i\int_{M}\left\{Tr_\g\{B\wedge F_A\}+Tr_\h\{C\wedge
  G_\Sigma\}\right\}}.
\end{equation}
The formal integration over $C$ and $B$ leads to
\begin{equation}
\label{eq_partcont2}
  Z = \int [\dd A ][\dd\Sigma]\, \delta( F_A)\, \delta ( G_\Sigma).
\end{equation}
Similarly to the treatment of $BF$-theory, this partition function is
then regularized on a triangulation $\Lambda$ of $M$. This amounts to
a translation from the differential picture~\cite{GiPf04} to the
integral picture~\cite{Pf03} of higher gauge theory. Similarly to
conventional gauge theory, the connection $A$ is discretized by
colouring the edges $e=(jk)\in\Lambda_1$ with group elements $g_e\in
G$. The connection $2$-form $\Sigma$ is in turn represented by group
elements $h_f\in H$ decorating the triangles
$f=(jk\ell)\in\Lambda_2$. We then recover the discretization described
in~\eqref{eq_triangle0} by reversing the procedure
of~\cite{GiPf04}. The vanishing fake curvature condition is
discretized on each triangle $f$ by replacing $\delta(F_A)$ by
\begin{equation}
  \delta_G\left(g_{ij}g_{jk}{\left(t(h_{ijk})g_{ik}\right)}^{-1}\right).
\end{equation}
The condition $\delta(G_\Sigma)$ on the curvature $3$-form for every
tetrahedron $T=(jk\ell m)\in\Lambda_3$ is turned into
\begin{equation}
\label{disccurvatures}
   \delta_H(h_{j\ell m}h_{jk\ell}(g_{jk}\rhd h_{k\ell m}^{-1})h_{jkm}^{-1}).
\end{equation}
The path integral measures of~\eqref{eq_partcont2} are discretized by replacing
\begin{eqnarray}
\label{measures}
  \int\dd A\quad&\mapsto&\quad \prod_{(jk)\in\Lambda_1}\int_G dg_{jk}\\
\label{measures2}
  \int\dd\Sigma\quad &\mapsto&\quad \prod_{(jk\ell)\in\Lambda_2}\int_H dh_{jk\ell},
\end{eqnarray}
where $dg_{jk}$ and $dh_{jk\ell}$ denote integration with respect to
the Haar measures of $G$ and $H$. By inserting~\eqref{disccurvatures},
\eqref{measures} and~\eqref{measures2} into \eqref{eq_partcont2}, we
obtain an expression proportional to~\eqref{eq_partition}.

It then turns out that this expression can be made independent of the
triangulation if one multiplies it by the appropriate `anomaly'
factors that render the expression equal to~\eqref{eq_partition}.

\acknowledgments

This research was supported in part by Perimeter Institute for Theoretical
Physics. The authors are grateful to Marco Mackaay and Timothy Porter for
correspondence and to Laurent Freidel and Jo{\~a}o Faria Martins for
discussions. We also thank the referee for valuable suggestions and for
spotting an error.

\appendix
%
\section{Pachner move invariance}
%
\label{app:pachner}

In the appendix, we give a self contained proof in terms of Pachner
moves that the partition function~\eqref{eq_partition} is independent
of the chosen triangulation and therefore well defined on equivalence
classes of combinatorial manifolds. By Whitehead's
theorem~\cite{Wh40}, it is thus even well defined on diffeomorphism
classes of smooth manifolds.

\subsection{Three-dimensional case}

We first sketch the proof of Pachner move invariance for the case $d=3$.

\subsubsection{The $1\leftrightarrow 4$ move}

We use the following notation. For every triangle $(jk\ell)\in\Lambda_2$, we
write
\begin{equation}
  g_{jk\ell} := t(h_{jk\ell})g_{jk}g_{k\ell}g_{j\ell}^{-1},
\end{equation}
and for every tetrahedron $(jk\ell m)\in\Lambda_3$,
\begin{equation}
  h_{jk\ell m} := h_{j\ell m}h_{jk\ell}(g_{jk}\rhd h_{k\ell m}^{-1})h_{jkm}^{-1}.
\end{equation}
Since the partition function~\eqref{eq_partition} is independent of the total
order of vertices, we need to verify the move only in one case. We denote the
vertices of the left hand side (one tetrahedron) by $1,2,3,4$ and the
additional vertex on the right hand side (four tetrahedra) by $5$. This
determines a total order, restricted to our subset of vertices. The partition
function on the two sides of the $1\leftrightarrow 4$ move then differs by the
following factors. On the l.h.s., we have the integrand
\begin{equation}
  \delta_H(h_{1234}),
\end{equation}
whereas on the r.h.s., we have integrals
\begin{equation}
  \int_{G^4}dg_{15}dg_{25}dg_{35}dg_{45}\,
  \int_{H^6}dh_{125}dh_{135}dh_{145}dh_{235}dh_{245}dh_{345}
\end{equation}
and the integrand
\begin{equation}
  \Biggl(\prod_{(jk\ell)\in M_2}\delta_G(g_{jk\ell})\Biggr)\,
    \Biggl(\prod_{(jk\ell m)\in M_3}\delta_H(h_{jk\ell m})\Biggr),
\end{equation}
where the products are over the following sets of simplices:
$M_2:=\{(125),(135),(145),(235),(245),(345)\}$ and
$M_3:=\{(1235),(1245),(1345),(2345)\}$. The numbers of the $k$-simplices on
both sides of the $1\leftrightarrow 4$ move are as follows (not taking into
account the remainder of the triangulation):

\begin{center}
\begin{tabular}{r|c|c|c|c}
&$|\Lambda_0|$&$|\Lambda_1|$&$|\Lambda_2|$&$|\Lambda_3|$\\
\hline
l.h.s. & 4&6&4&1\\
r.h.s. & 5&10&10&4\\
\hline
\end{tabular}
\end{center}

In order to verify the $1\leftrightarrow 4$ move, we consider the r.h.s\ and
first integrate over $g_{15}$, exploiting $\delta_G(g_{125})$, \ie\ the
integral over $g_{15}$ and the integrand $\delta_G(g_{125})$ both disappear,
and all other occurrences of $g_{15}$ in the integrand are replaced by
$t(h_{125})g_{12}g_{25}$. We then integrate over $g_{25}$, exploiting
$\delta_G(g_{235})$, and over $g_{35}$, exploiting $\delta_G(g_{345})$. At
this stage, the integral over $g_{45}$ is trivial, \ie\ over a constant
integrand.

Finally, we integrate over $h_{135}$, exploiting $\delta_H(h_{1345})$, \ie\
substituting $h_{135}=h_{145}h_{134}(g_{13}\rhd h_{345}^{-1})$ everywhere else
in the integrand, and we integrate over $h_{125}$, exploiting
$\delta_H(h_{1245})$, and over $h_{235}$, exploiting $\delta_H(h_{2345})$. One
can now show that the remaining integrand of the r.h.s.\ equals
\begin{equation}
  {(\delta_G(e))}^3\delta_H(h_{1234}) = {|G|}^3\delta_H(h_{1234}),
\end{equation}
and so the remaining three integrals over $h_{145}$, $h_{245}$, and $h_{345}$
are trivial. In order to show this, we make use the condition
\begin{equation}
\label{eq_triangle}
  g_{j\ell} = t(h_{jk\ell})g_{jk}g_{k\ell}
\end{equation}
for $(jk\ell)\in\{(123),(124),(234)\}$. This is possible because these
triangles are present on both sides of the move, and so the
corresponding $\delta_G(g_{jk\ell})$ that enforces the
condition~\eqref{eq_triangle}, are part of the integrand. Finally, the
prefactor
${|G|}^{-|\Lambda_0|+|\Lambda_1|-|\Lambda_2|}{|H|}^{|\Lambda_0|-|\Lambda_1|+|\Lambda_2|-|\Lambda_3|}$
is ${|G|}^{-2}{|H|}^1$ on the l.h.s.\ and ${|G|}^{-5}{|H|}^1$ on the
r.h.s., compensating for the ${|G|}^3$ from the left over $\delta_G$ of
the integrand.

\subsubsection{The $2\leftrightarrow 3$ move}

The numbers of $k$-simplices on the two sides of the $2\leftrightarrow 3$ move
are as follows:

\begin{center}
\begin{tabular}{r|c|c|c|c}
&$|\Lambda_0|$&$|\Lambda_1|$&$|\Lambda_2|$&$|\Lambda_3|$\\
\hline
l.h.s. & 5&9&7&2\\
r.h.s. & 5&10&9&3\\
\hline
\end{tabular}
\end{center}

We order the vertices in such a way that the l.h.s.\ has the tetrahedra
$(1234)$ and $(2345)$, sharing the triangle $(234)$, whereas the r.h.s\ has
the tetrahedra $(1235)$, $(1245)$ and $(1345)$, all sharing the edge $(15)$
and each two of them sharing one of the triangles $(125)$, $(135)$ and
$(145)$.

On the l.h.s.\ of the $2\leftrightarrow 3$ move, we therefore have the
integral
\begin{equation}
  \int_Hdh_{234}
\end{equation}
and the integrand
\begin{equation}
  \delta_G(g_{234})\,\delta_H(h_{1234})\delta_H(h_{2345}),
\end{equation}
whereas on the r.h.s\ we have the integrals
\begin{equation}
  \int_Gdg_{15}\int_{H^3}dh_{125}dh_{135}dh_{145}
\end{equation}
and the integrand
\begin{equation}
  \delta_G(g_{125})\delta_G(g_{135})\delta_G(g_{145})\,
  \delta_H(h_{1235})\delta_H(h_{1245})\delta_H(h_{1345}).
\end{equation}
All other integrals and all other factors of the integrand are the same on
both sides of the move.

In order to simplify the l.h.s., we integrate over $h_{234}$, exploiting
$\delta_H(h_{2345})$. In the remaining integrand, we therefore substitute
$h_{234}=h_{245}^{-1}h_{235}(g_{23}\rhd h_{345})$. The integrand of the
l.h.s.\ thus reduces to
\begin{equation}
  \delta_G(e)\delta_H(h_{124}(g_{12}\rhd(h_{245}^{-1}h_{235}))h_{123}^{-1}(g_{13}\rhd h_{345})h_{134}^{-1}).
\end{equation}
In order to simplify the r.h.s., we integrate over $g_{15}$, exploiting
$\delta_G(g_{135})$, over $h_{125}$, exploiting $\delta_H(h_{1235})$, and over
$h_{135}$, exploiting $\delta_H(h_{1345})$. The remaining integral over
$h_{145}$ turns out to be trivial if one uses~\eqref{eq_triangle} for all
$(jk\ell)\in\{(123),(235),(134),(345)\}$. The integrand of the r.h.s\ reduces
to
\begin{equation}
  {(\delta_G(e))}^2\delta_H(h_{124}(g_{12}\rhd(h_{245}^{-1}h_{235}))h_{123}^{-1}(g_{13}\rhd h_{345})h_{134}^{-1}).
\end{equation}
Again, the different powers of $\delta_G(e)=|G|$ are compensated for by the
prefactors. These are ${|G|}^{-3}{|H|}^1$ on the l.h.s.\ and
${|G|}^{-4}{|H|}^1$ on the r.h.s.

\subsection{Four-dimensional case}

We now sketch the proof of Pachner move invariance for the case $d=4$.

\subsubsection{The $1\leftrightarrow 5$ move}

The numbers of the $k$-simplices on both sides of the move are as follows:

\begin{center}
\begin{tabular}{r|c|c|c|c|c}
&$|\Lambda_0|$&$|\Lambda_1|$&$|\Lambda_2|$&$|\Lambda_3|$&$|\Lambda_4|$\\
\hline
l.h.s. & 5&10&10&5&1\\
r.h.s. & 6&15&20&15&5\\
\hline
\end{tabular}
\end{center}

Again, since the partition function~\eqref{eq_partition} does not depend on
the total order of the vertices, we need to show this move only for one case.
We order the vertices such that the l.h.s\ consists of the $4$-simplex
$(23456)$ whereas the r.h.s.\ contains the five $4$-simplices $(13456)$,
$(12456)$, $(12356)$, $(12346)$ and $(12345)$. On the r.h.s., we therefore
have the triangles $(jk\ell)\in
M_2:=\{(123),(124),(125),(126),(134),(135),(136),(145),(146),(156)\}$ and the
edges $(jk)\in M_1:=\{(12),(13),(14),(15),(16)\}$.

In order to compare the l.h.s\ with the r.h.s.\ of the $1\leftrightarrow 5$
move, we have to show that the integrals
\begin{equation}
  \int_{G^5}\prod_{(jk)\in M_1}dg_{jk}\,
  \int_{H^{10}}\prod_{(jk\ell)\in M_2}dh_{jk\ell}
\end{equation}
and the integrand
\begin{equation}
  \Biggl(\prod_{(jk\ell)\in M_2}\delta_G(g_{jk\ell})\Biggr)\,
  \Biggl(\prod_{(jk\ell m)\in M_3}\delta_H(h_{jk\ell m})\Biggr)
\end{equation}
on the r.h.s.\ reduce to $1$. Here,
$M_3:=\{(1234),(1235),(1236),(1245),(1246),(1256),(1345),(1346),(1356),(1456)\}$.

In order to simplify the r.h.s., we integrate over $h_{123}$, exploiting
$\delta_H(h_{1234})$, and over $g_{12}$, exploiting $\delta_G(g_{123})$, and
make use of~\eqref{eq_triangle} for $(jk\ell)=(234)$. We then integrate over
$h_{124}$, exploiting $\delta_H(h_{1245})$, and over $g_{13}$, exploiting
$\delta_G(g_{124})$, and make use of~\eqref{eq_triangle} for
$(jk\ell)\in\{(234),(235)\}$. We then integrate over $h_{125}$, exploiting
$\delta_H(h_{1236})$, and over $g_{14}$, exploiting $\delta_G(g_{125})$, and
make use of~\eqref{eq_triangle} for $(jk\ell)=(236)$.

Finally, we integrate over $g_{15}$, exploiting $\delta_G(g_{146})$, and over
$h_{134}$, exploiting $\delta_H(h_{1346})$, over $h_{135}$, exploiting
$\delta_H(h_{1356})$, and over $h_{145}$, exploiting $\delta_H(h_{1456})$. We
make use of~\eqref{eq_triangle} for $(jk\ell)=\{(234),(346),(356),(456)\}$. We
then use the condition that
\begin{equation}
\label{eq_tetrahedron}
  h_{jkm} = h_{j\ell m}h_{jk\ell}(g_{jk}\rhd h_{k\ell m}^{-1})
\end{equation}
for all $(jk\ell m)\in\{(2345),(2346),(2356),(3456)\}$. The remaining
integrals over $h_{126}$, $h_{136}$, $h_{146}$ and $h_{156}$ are trivial as
well as that over $g_{16}$. The integrand reduces to
\begin{equation}
  {(\delta_G(e))}^6{(\delta_H(e))}^3={|G|}^6{|H|}^4.
\end{equation}
The prefactor
${|G|}^{-|\Lambda_0|+|\Lambda_1|-|\Lambda_2|}{|H|}^{|\Lambda_0|-|\Lambda_1|+|\Lambda_2|-|\Lambda_3|}$
equals ${|G|}^{-5}{|H|}^0$ on the l.h.s.\ and ${|G|}^{-11}{|H|}^{-4}$ on the
r.h.s\ and therefore compensates for these left-over factors.

\subsubsection{The $2\leftrightarrow 4$ move}

The numbers of the $k$-simplices on both sides of the move are as follows:

\begin{center}
\begin{tabular}{r|c|c|c|c|c}
&$|\Lambda_0|$&$|\Lambda_1|$&$|\Lambda_2|$&$|\Lambda_3|$&$|\Lambda_4|$\\
\hline
l.h.s. & 6&14&16&9&2\\
r.h.s. & 6&15&20&14&4\\
\hline
\end{tabular}
\end{center}

We order the vertices in such a way that on the l.h.s., we have the
$4$-simplices $(23456)$ and $(13456)$ whereas on the r.h.s., there are
$(12456)$, $(12356)$, $(12346)$ and $(12345)$. On the l.h.s., there is one
tetrahedron $(3456)$ whereas on the r.h.s., there are six, namely $(1234)$,
$(1235)$, $(1236)$, $(1245)$, $(1246)$ and $(1256)$. All other tetrahedra are
part of the common boundary of both sides of the move. On the r.h.s., we also
have the triangles $(123)$, $(124)$, $(125)$ and $(126)$ and the edge $(12)$.

The integrals and factors of the integrand that differ on both sides of the
$2\leftrightarrow 4$ move are as follows. On the l.h.s., there is the
integrand
\begin{equation}
  \delta_H(h_{3456})
\end{equation}
whereas on the r.h.s., we have the integrals
\begin{equation}
  \int_Gdg_{12}\,\int_{H^4}dh_{123}dh_{124}dh_{125}dh_{126}
\end{equation}
and the integrand
\begin{equation}
  \Biggl(\prod_{(jk\ell)\in M_2}\delta_G(g_{jk\ell})\Biggr)\,
  \Biggl(\prod_{(jk\ell m)\in M_3}\delta_H(h_{jk\ell m})\Biggr),
\end{equation}
where $M_2:=\{(123),(124),(125),(126)\}$ and
$M_3:=\{(1234),(1235),(1236),(1245),(1246),(1256)\}$.

The l.h.s.\ simplifies to $\delta_H(e)=|H|$ because of the following general
result.

\begin{lemma}
Given a $4$-simplex $(jk\ell mn)$ with a colouring that
satisfies~\eqref{eq_tetrahedron} for four of the tetrahedra $(k\ell mn)$,
$(j\ell mn)$, $(jkmn)$ and $(jk\ell n)$ and~\eqref{eq_triangle} for all
triangles in their boundary, then~\eqref{eq_tetrahedron} also holds on the
fifth tetrahedron $(jk\ell m)$.
\end{lemma}

\begin{proof}
Consider $h_{jk\ell m}=h_{j\ell m}h_{jk\ell}(g_{jk}\rhd h_{k\ell
m}^{-1})h_{jkm}^{-1}$ and use the condition~\eqref{eq_triangle} for
$(jk\ell)$, $(jkm)$, $(jkn)$, $(j\ell m)$, $(j\ell n)$, $(jmn)$, $(k\ell m)$
and $(k\ell n)$. This implies $h_{jk\ell m}=e$.
\end{proof}

In order to simplify the r.h.s., we integrate over $h_{124}$, exploiting
$\delta_H(h_{1234})$, over $h_{123}$, exploiting $\delta_H(h_{1235})$, and
over $h_{125}$, exploiting $\delta_H(h_{h_1256})$. Use the
condition~\eqref{eq_triangle} for all
$(jk\ell)\in\{(134),(234),(135),(235),(156),(256)\}$. Then we integrate over
$g_{12}$, exploiting $\delta_G(g_{126})$ and use~\eqref{eq_triangle} for
$(jk\ell)\in\{(156),(256)\}$ again. The last remaining integral over $h_{126}$
is then trivial, and the integrand reduces to
\begin{equation}
  {(\delta_G(e))}^3{(\delta_H(e))}^3={|G|}^3{|H|}^3.
\end{equation}
The difference in powers of $|G|$ and $|H|$ is compensated for by the
prefactors which equal ${|G|}^{-8}{|H|}^{-1}$ on the l.h.s.\ and
${|G|}^{-11}{|H|}^{-3}$ on the r.h.s.

\subsubsection{The $3\leftrightarrow 3$ move}

We order the vertices in such a way that one the l.h.s.\ of the
$3\leftrightarrow 3$ move, we have the $4$-simplices $(23456)$, $(13456)$ and
$(12456)$ whereas on the r.h.s.\ they are $(12356)$, $(12346)$ and $(12345)$.
Six tetrahedra therefore form the common boundary of both sides of the move
whereas on each side there are three tetrahedra shared by two $4$-simplices.
On the l.h.s.\ these are $(1456)$, $(2456)$ and $(3456)$ and on the r.h.s.\
$(1234)$, $(1235)$ and $(1236)$. On the l.h.s\ we therefore have the triangle
$(456)$ and on the r.h.s\ $(123)$. All other triangles appear on both sides of
the move.

The integral and integrand for the l.h.s.\ read
\begin{equation}
  \int_Hdh_{456}\delta_G(g_{456})\,\delta_H(h_{3456})\delta_H(h_{2456})\delta_H(h_{1456}),
\end{equation}
whereas for the r.h.s.\ we have
\begin{equation}
  \int_Hdh_{123}\delta_G(g_{123})\,\delta_H(h_{1234})\delta_H(h_{1235})\delta_H(h_{1236}).
\end{equation}
In order to simplify the l.h.s., we make use of the fact that $\delta_H(-)$ is
constant on the orbits of $G$ on $H$ and also on the conjugacy classes, \ie\
\begin{equation}
  \delta_H(h_{3456}) = \delta_H((g_{23}^{-1}\rhd h_{346}^{-1})
  (g_{23}^{-1}\rhd h_{356})(g_{23}^{-1}\rhd h_{345})h_{456}^{-1}).
\end{equation}
We then integrate over $h_{456}$, exploiting $\delta_H(h_{3456})$. The
integrand reduces to
\begin{equation}
\label{eq_ttintegrand}
  \delta_G(e){(\delta_H(e))}^2=|G|{|H|}^2
\end{equation}
if we make use of the condition~\eqref{eq_triangle} for all
$(jk\ell)\in\{(134),(234),(345),(346),(356)\}$ and of~\eqref{eq_tetrahedron}
for all $(jk\ell m)\in\{(1345),(1346),(1356),(2345),(2346),(2356)\}$.

In order to simplify the r.h.s., we integrate over $h_{123}$, exploiting
$\delta_H(h_{1234})$. The integrand reduces to~\eqref{eq_ttintegrand}, too, if
we make use of~\eqref{eq_triangle} for
$(jk\ell)\in\{(123),(124),(134),(234)\}$ and of~\eqref{eq_tetrahedron} for
$(jk\ell m)\in\{(1245),(1246),(1345),(1346),(2345),(2346)\}$. The numbers of
$k$-simplices agree on both sides of the $3\leftrightarrow 3$ move for all
$k$, and the prefactors play no role in this case.

%
\section{State sum models with $2$-categories}
%
\label{app:mackaay}

Under certain conditions, the partition function~\eqref{eq_partition}
for $d=4$ forms a special case of Mackaay's state
sum~\cite{Ma99}. Here we explain in detail in which case this happens
and which assumptions of~\cite{Ma99} are violated in more general
situations.

First, we follow~\cite{Ma00} in which Mackaay specializes his state
sum of~\cite{Ma99} to the case of finite groups, and describe the
common special case with our model~\eqref{eq_partition}.

Recall that weak $2$-groups~\cite{BaLa04} are algebraic models for
pointed and connected homotopy $2$-types as follows. Given any path
connected CW-complex $X$ with $1$-skeleton $X_1$ and base point $p\in
X_1\subseteq X$, there is a weak $2$-group $\Pi_2(X,X_1,p)$ defined as
follows. It has only a single object $p$. The $1$-morphisms are the
continuous closed curves in $X_1$ with base point $p$. The
$2$-morphisms are bigon-shaped surfaces between two such curves,
continuously mapped into $X$, up to homotopy. It can be shown that
$\Pi_2(X,X_1,p)$ forms a bicategory with one object. Furthermore, it
turns out that each $2$-morphism has got a vertical inverse and that
each $1$-morphism has got an inverse up to $2$-isomorphism, and so
$\Pi_2(X,X_1,p)$ forms a weak $2$-group as defined in~\cite{BaLa04}.

Homotopy equivalent based pairs of spaces $(X,X_1,p)\simeq (Y,Y_1,q)$
yield $\Pi_2(X,X_1,p)$ and $\Pi_2(Y,Y_1,q)$ that are equivalent as
weak $2$-groups. There are two characterizations of $\Pi_2(X,X_1,p)$
up to equivalence of weak $2$-groups that are relevant in the
following.

First, by the coherence theorem for bicategories, $\Pi_2(X,X_1,p)$ is
equivalent to a (strict) $2$-category. It can be shown that it is even
equivalent to a strict $2$-group and can thus be characterized by a
crossed module $(G,H,\rhd,t)$ of groups. In this case, we have
$G\cong\pi_1(X_1)$ and $H\cong\pi_2(X,X_1)$. Note that this is in
general a non-abelian group. The action $\rhd$ is the action of
$\pi_1(X_1)$ on $\pi_2(X,X_1)$ by the change of base point, and
$t\colon\pi_2(X,X_1)\to\pi_1(X_1)$ is the restriction to the
boundary. This way, the crossed module $(G,H,\rhd,t)$ is determined up
to equivalence in the $2$-category of crossed modules.

Second, every bicategory with one object forms a weak monoidal
category, and any weak monoidal category is equivalent (as a weak
monoidal category) to any of its skeleta. This result can be extended
to weak $2$-groups whose skeleta are precisely the \emph{special
$2$-groups} of~\cite{BaLa04}. Passing to a skeleton in this way
amounts to characterizing $\Pi_2(X,X_1,p)$ in terms of its Postnikov
data $(K,A,\blacktriangleright$,$\alpha$) where $A:=\pi_2(X)$
(abelian), $K:=\pi_1(X)$, $\blacktriangleright$ denotes the action of
$\pi_1(X)$ on $\pi_2(X)$ by the change of base point, $\alpha$ is the
Postnikov $k$-invariant which is an $A$-valued algebraic $3$-cocycle
on $K$.  Starting from the crossed module $(G,H,\rhd,t)$, the groups
$A$ and $K$ appear when one extends the map $t$ to a $4$-term exact
sequence of groups,
\begin{equation}
\xymatrix{
  \{0\}\ar[r]&
  A\ar@{^{(}->}[r]^{\imath}&
  H\ar[r]^{t}&
  G\ar@{->>}[r]^{\pi}&
  K\ar[r]&
  \{e\}
}
\end{equation}
\ie\ $A\cong\ker(t)$ and $K\cong\coker(t)\cong G/t(H)$. The action
$\rhd$ of $G$ on $H$ by automorphisms induces an action $\blacktriangleright$
of $K$ on $A$ by automorphisms. The Postnikov $k$-invariant can then be
constructed from a section of the map $\pi$ (using, in general, the axiom of
choice) and a diagram chase. Mackaay's $G$ and $H$ in~\cite{Ma00} are our $K$
and $A$, respectively.

A close look at~\cite{Ma00} reveals that in the case in which the
action $\blacktriangleright$ of $K$ on $A$ is trivial, our partition
function~\eqref{eq_partition} agrees with Mackaay's state sum
of~\cite{Ma00}. The case in which $K$ acts trivially on $A$, however,
is far from generic. For example, let $(G,H,\rhd,t)$ be a crossed
module of groups in which $H$ is abelian, $t(h)=e\in G$ for all
$h\in H$ and in which $G$ acts non-trivially on $H$. In this case,
$A\cong H$, $K\cong G$, and so $K$ acts non-trivially on $A$. There
exist many examples of this type. As soon as $K$ acts non-trivially on
$A$, however, Mackaay's construction~\cite{Ma00} is no longer a
special case of his state sum~\cite{Ma99}. For the general case, we
cannot use~\cite{Ma00}, but rather have to go back to the state sum as
defined in~\cite{Ma99}.

Let us explain how to define for every crossed module of groups
$(G,H,\rhd,t)$ a semi-strict monoidal $2$-category with duals in such
a way that Mackaay's state sum~\cite{Ma99} agrees with our partition
function~\eqref{eq_partition}. This $2$-category is semi-strict and
pivotal, but not spherical, and so the proof of Pachner move
invariance used in~\cite{Ma99} no longer applies. The fact
that~\eqref{eq_partition} is nevertheless Pachner move invariant as
was known from~\cite{Po96,Po98} and as we have confirmed in
Appendix~\ref{app:pachner} above, suggests that one ought to
generalize the definition of \emph{spherical} and modify the proof of
Pachner move invariance in~\cite{Ma99} accordingly in order to
encompass our example~\eqref{eq_partition} as well.

The $2$-group associated with the crossed module $(G,H,\rhd,t)$ forms
a small category whose objects are elements of $G$ and whose
$1$-morphisms $f\colon g_1\to g_2$ are elements $f=(h,g)\in H\rtimes
G$ that satisfy $g=g_1$ and $t(h)g=g_2$. The composition of
$1$-morphisms is the vertical composition. The $2$-category to
consider is the discrete $2$-category on this small category, \ie\ its
objects are elements of $G$, its $1$-morphisms $f\colon g_1\to g_2$
are elements $f=(h,g)\in H\rtimes G$ as above, and for every
$1$-morphism $f$, there is only the identity $2$-morphism.

This $2$-category has a semi-strict monoidal structure as follows. For
objects $g_1,g_2\in G$, we have $g_1\otimes g_2=g_1g_2$. For
$1$-morphisms $f=(h,g_1)\colon g_1\to g_2$ we have $\tilde g\otimes
f=(e,\tilde g)\cdot(h,g_1)=(\tilde g\rhd h,\tilde gg_1)$ and
$f\otimes\tilde g=(h,g_1)\cdot (e,\tilde g)=(h,g_1\tilde g)$. The
discreteness of the $2$-category determines the monoidal structure on
$2$-morphisms. The monoidal unit $\1=e\in G$ is the unit of $G$.

For $1$-morphisms $(h_1,g_1)\colon g_1\to g_1^\prime$ and
$(h_2,g_2)\colon g_2\to g_2^\prime$, the tensorator $2$-isomorphism
is the identity $2$-morphism associated with the following equality
of $1$-morphisms:
\begin{equation}
  ((h_1,g_1)\otimes g_2^\prime)\circ (g_1\otimes (h_2,g_2))
  = (g_1^\prime\otimes (h_2,g_2))\circ((h_1,g_1)\otimes g_2).
\end{equation}
Duality is defined as follows. The dual of an object $g\in G$ is its
inverse $g^\ast=g^{-1}$ with unit and counit $\imath_g=(e,e)\colon\1\to
g\otimes g^\ast$ and $e_g=(e,e)\colon g^\ast\otimes g\to\1$. The
triangulator is the identity $2$-morphism associated with the
following equality of $1$-morphisms:
\begin{equation}
  (e_g\otimes g)\circ (g\otimes\imath_g)=\id_g.
\end{equation}
The dual of a $1$-morphism $(h,g_1)\colon g_1\to g_2$ is its vertical
inverse ${(h,g_1)}^\ast = (h^{-1},g_2)$. The unit and counit for this
dual as well as the duals of $2$-morphisms are already determined
because of discreteness of the $2$-category.

For each $1$-morphism $(h,g)\colon g_1\to g_2$, the $1$-morphisms
${}^\sharp f$ and $f^\sharp$ of~\cite{Ma99} turn out to be ${}^\sharp
f=(g_2^{-1}\rhd h,g_2^{-1})\colon g_2^\ast\to g_1^\ast$ and
$f^\sharp=(g_1^{-1}\rhd h,g_2^{-1})\colon g_2^\ast\to g_1^\ast$. And
so the semi-strict monoidal $2$-category is pivotal because of the
identity $2$-morphism associated with the equality of $1$-morphisms
$f^\sharp = {}^\sharp f$.

As remarked in~\cite{Ma04}, however, left- and right-traces of a
$1$-morphism $f=(h,g)\colon g\to t(h)g$ turn out to be
$\tr_L(f)=(g^{-1}\rhd h,e)$ and $\tr_R(f)=(h,e)$, respectively. Unless
$G$ acts trivially on $H$, these are in general distinct
$1$-morphisms, and so in the discrete $2$-category, there cannot exist
any $2$-isomorphism between them. Therefore, the $2$-category is not
spherical. Nevertheless, the state sum defined in~\cite{Ma99} agrees
with~\eqref{eq_partition} for $d=4$ for any finite crossed module
$(G,H,\rhd,t)$, up to an overall prefactor
${|G|}^{-|\Lambda_0|}{|H|}^{|\Lambda_0|-|\Lambda_1|}$.

\newenvironment{hpabstract}{%
  \renewcommand{\baselinestretch}{0.2}
  \begin{footnotesize}%
}{\end{footnotesize}}%
\newcommand{\hpeprint}[2]{%
  \href{http://www.arxiv.org/abs/#1}{\texttt{arxiv:#1#2}}}%
\newcommand{\hpspires}[1]{%
  \href{http://www.slac.stanford.edu/spires/find/hep/www?#1}{SPIRES Link}}%
\newcommand{\hpmathsci}[1]{%
  \href{http://www.ams.org/mathscinet-getitem?mr=#1}{\texttt{MR #1}}}%
\newcommand{\hpdoi}[1]{%
  \href{http://dx.doi.org/#1}{\ Journal Link}}%

\end{document}